# Impact of a non-deterministic rekey interval on the performance of arc4random

Keelan Cannoo
Student in Bsc Computer Science
University of Mauritius
+230 5929 4259
keelan.cannoo@umail.uom.ac.mu

Jeevesh Rishi Dindyal
Student in Bsc Computer Science
University of Mauritius
+230 5719 8900
jeevesh.dindyal@umail.uom.ac.mu

## ABSTRACT
This paper is intended to provide the readers with a clear understanding of the modification done to the pseudo-random number generator (PRNG) arc4random. The arc4random PRNG was modified to implement a non-deterministic rekey interval. A comparison was made with the unpatched version regarding performance and the rekey interval's randomization. All measurements were done when the rekey was performed. A slight increase in performance was observed, and the rekey interval was randomized as expected.

## CCS Concepts
•Security and privacy~Cryptography~Information-theoretic techniques

## Keywords

## 1. INTRODUCTION

Arc4random is a pseudo-random number generator first introduced in the OpenBSD 2.1 project. Arc4random was first implemented using the RC4 algorithm; however, due to vulnerability in RC4 and considered insecure[7], it was later updated with ChaCha20. The ChaCha20 algorithm replaced the RC4 in the OpenBSD 5.5 project. Arc4random is encouraged for random number generators as others are deficient in quality, portability, standardization, or availability. The arc4random will return a 32-bit value successfully. The arc4random does not have any reserve number to identify error as it always return successfully there it has a range of $2^{31} - 1$[6]. It is difficult to predict the outcome of the arc4random function, but the interval of the rekey was predictable as a constant was used to determine when to perform a rekey. Therefore, to improve the security of the PRNG, the rekey interval was randomized and the consequences on its performance were asessed. Pseudo-random numbers play a significant role in computing and communication. Some applications of arc4random are described below:

LibreSSL uses arc4random[2]

- to generate the client random, the server random and the pre master key necessary during TLS handshake.
- to generate random keys and key name for RFC4507 tickets
- to generate session ID

Unbound DNS[5] uses arc4random to

- pick random IP
- The certificate RFC7958 update is done by fetching root-anchors.xml and root-anchors.p7s via SSL. The HTTPS certificate can be logged but is not validated (HTTPS for channel security; the security comes from the certificate). The 'data.iana.org' domain name A and AAAA is resolved without DNSSEC. It tries a random IP until the transfer succeeds. It then checks the p7s signature.
- debug programs
    - delayer - This program delay queries made. It performs as a proxy for another server and delays queries to it.
    - streamtcp – This program performs multiple DNS queries on TCP

Libevent[4]

- evdns

opencoff/portable-lib[3] uses arc4random to

- generate a random salt for the seeded hash functions

openNTPD[1]

- client query sends a random 64-bit number as transmit time

## 2. Implementation & Testing

Our goal is to prevent the prediction of the rekey interval by using the ChaCha20 algorithm and giving a random value. The ChaCha20 algorithm generates a stream of random values. A constant is mod with the algorithm's output and adds up to the same constant value[8].

```
/*rs->rs_count = 1600000;*/

    /* rekey interval should not be predictable */
    chacha_encrypt_bytes(&rsx->rs_chacha, (uint8_t *)&rekey_fuzz,
  (uint8_t *)&rekey_fuzz, sizeof(rekey_fuzz));
    rs->rs_count = REKEY_BASE + (rekey_fuzz % REKEY_BASE);
```

*Drawing 1: snippet of code implemented*

## 2.1 Raw evaluation of Arc4random

### 2.1.1 Generation time test

A program that generates 151 MB worth of pseudo-random numbers (39600000 integers) is run. The time it takes to complete and the number of times rekey happens is recorded. This process is repeated for 10 set of values, and the mean is calculated. The test is done for the patched and unpatched versions of arc4random, and the measurements are compared.

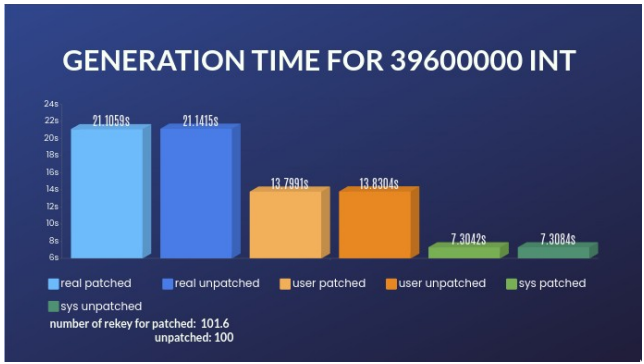

*Figure 1: Comparison of generation time performance*

Below is the performance increase and reduction in time calculate in percentage:

|  | reduction in time | increase in performance |
|---|---|---|
| real | 0.16% | 0.17% |
| user | 0.22% | 0.23% |
| sys | 0.057% | 0.056% |

### 2.1.2 Chi-square test for quality of randomness

A program generates 39600000 pseudo-random integers between 0 and 99 inclusive using arc4random_uniform. The number of rekey occurrences is also recorded. Arc4random did 100 rekeying for the unpatched version, and for the patched version, it ranged from 90 to 100.The statistical chi-square goodness of fit test is performed on the generated numbers. (A random number is equally probable everywhere). Patched and unpatched versions are tested. The result is summarized below:

|  | patched | unpatched | difference (%) |
|---|---|---|---|
| chi square | 6.829 | 7.261 | 5.955 |

### 2.1.3 LibreSSL

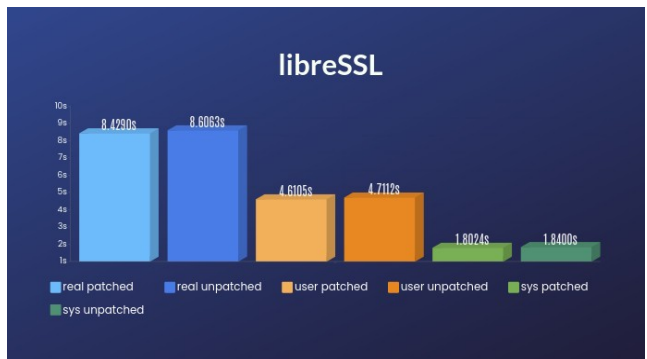

*Figure 2: Comparison of libreSSL performance*

A TLS server is set up, and a client repeatedly attempts to connect. Each time the client connects to the server, a TLS handshake occurs during which random numbers are generated (for client random, server random, and pre-master). The time to establish 500 requests is recorded for the patched and unpatched versions of LibreSSL. The measurements are shown below:

Below is the performance increase and reduction in time expressed in percentage:

|  | reduction in time | increase in performance |
|---|---|---|
| real | 2.1% | 2.1% |
| user | 2.1% | 2.2% |
| sys | 2.0% | 2.1% |

### 2.1.4 Libevent

The libevent library is used by Tor Browser. The time taken to launch the web browser is recorded. The test is repeated for 1000 values and the mean is calculated for both the patched and unpatched versions. The measurements are shown below.

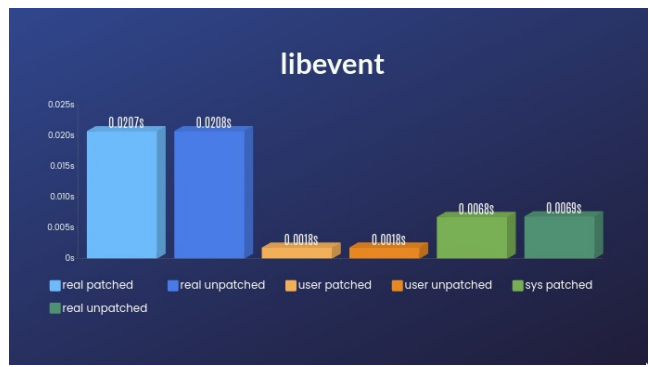

*Figure 3: Comparison of libevent performance*

Below are the performance increase and reduction in time calculated in percentage:

|  | reduction in time | increase in performance |
|---|---|---|
| real | 0.16% | 0.16% |
| user | 0.37% | 0.37% |
| sys | 1.42% | 1.44% |

### 2.1.5 Opencoff portable-lib

For the opencoff portable-lib, as it is a portable library for programmers to include in their project, a function call was done on the arc4random function. The set of 1000 calls was done, and the average run time of arc4random was computed for both the patched and unpatched versions.

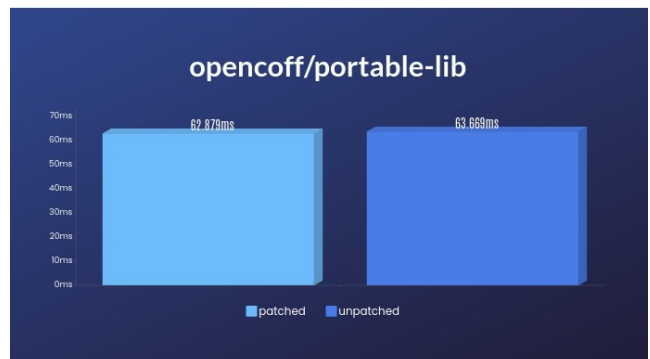

*Figure 4: Comparison of opencoff portable-lib performance*

|  | eduction in time | increase in performance |
|---|---|---|
| patched | 1.2% | 1.3% |

### 2.1.6 OpenNTPD

In the OpenNTPD package, a clock syncing was made with the two servers 2.arch.pool.ntp.org and time.cloudflare.com without constraint to be able to call a function named query_client, which uses arc4random. The time taken to make a rekey was recorded and repeated ten times for the patched and unpatched versions.

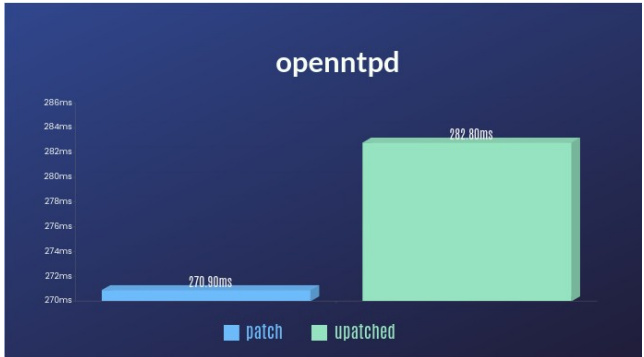

Figure 5: Comparison of openntpd performance

|         | reduction in time | increase in performance |
|---------|-------------------|-------------------------|
| patched | 4.4%              | 4.2%                    |

### 2.1.7 Unbound DNS

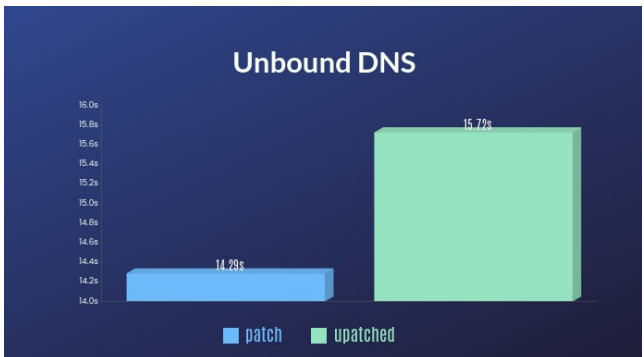

Figure 6: Comparison of Unbound DNS performance

The unbound DNS server was tested, using the command Unbound-host to connect to the dns "www.example.org", which uses the Unbound validating resolver to query for the hostname and display results. Every query for the hostname call arc4random and rekey interval was changed. The test was made 100 times with a stable connection of 20mbps. The values were computed for both patched and unpatched versions and compared.

|         | reduction in time | increase in performance |
|---------|-------------------|-------------------------|
| Patched | 9.1%              | 10%                     |

## 3. CONCLUSION

To conclude, the patched version shows a slight increase in the performance on all tests. The randomness of the rekey interval performed as expected. The rekey interval change at random values instead of the constant rekey interval as before. The patched version did not cause any harm to any other unit of the libraries tested. Every library performed their task as expected.

## 4. ACKNOWLEDGMENTS

We would like to express our gratitude to Mr Loganaden Velvindron for helping us and guiding us on the patching and testing of the libraries. Moreover, we thank the cyberstorm.mu group and specially one of his member, Mr Alex Bissessur. Finally we thank Mr Damien Miller the original author of the code we have been working on and testing.